# Unconventional optical microcavities hosting multiple exceptional points

Arnab Laha, Somnath Ghosh *

Institute of Radio Physics and Electronics, University of Calcutta, Kolkata-700009, India
Publications Department, OSA—The Optical Society, 2010 Massachusetts Avenue N.W., Washington, D.C. 20036
*Corresponding author: somiit@rediffmail.com

**Recently, presence of hidden singularities known as exceptional points (EPs) in non-Hermitian quantum systems has opened up a tremendous interest in different domains of physics owing to their unique unconventional physical effects. Effectively for such systems an EP appears as a fascinating topological defect where two mutually interacting eigenstates of the system coalesce. In this work, we report occurrence of EP via avoided crossing between coupled resonance states in an optical microcavity with spatially varying gain loss profile, an optical analogue of non-Hermitian system. With suitably tailored system openness and coupling strength, internally coupled resonances can exhibit EP in a medium with simultaneous presence of loss and gain. We explore the characteristics behaviors of energy eigenvalues in complex energy plane and corresponding eigenstates when the control parameters of the system are adiabatically changed, such a way that they encircle the EP. In this letter, we first ever exploit the above scheme to analyze the optical performance in terms of stability and rigidity of a microcavity simultaneously hosting such multiple EPs.**

Virtually for all domains of physics, the dependence on parameters of the energies and widths of resonance states has always been a great physical insight. In that direction, the repulsion of levels in the complex energy plane is a very interesting phenomenon. This level repulsion in the complex energy plane can appear as a crossing of their real parts with simultaneous avoided crossing of corresponding imaginary parts or vice versa. These aspects have been widely discussed in different contexts of electromagnetic resonators [1, 2], optical microcavity and laser absorber like various absorptive media in solid state physics [3-5]. The relation of these types of crossing and anti-crossing can be understood by proper positioning the particular singular points of the spectrum that can be described as the coalescence of two eigenstates, i.e. two interacting resonance states are very close to degeneracy [6]. While the well-known properties associated with a genuine degeneracy of Hermitian operators are no longer valid. These singularities have been coined as exceptional points (EPs) by T. Kato [7]. Generally, EPs are associated with symmetry breaking for PT-symmetric Hamiltonian. For suitable description of EP, the parameter space has to be two dimensional, i.e. there exists at least two real parameters and the resonances should appear as complex eigenvalues whose real part correspond to energy and imaginary part correspond to resonance width. The appearance of exceptional points, originating from a parameter dependent eigenvalue problem, and its dramatic effect in non-Hermitian open quantum system have theoretically been shown in laser field [8, 9], optical waveguides [10, 11], resonators [12] and microwave cavity[13]. In quantum systems their existence has been proved in atomic [14–15] and molecular [16] spectra, in the scattering of particles at potential barriers [17], in atom waves [18-19], and in non-Hermitian Bose-Hubbard models [20]. Tremendous efforts have been put forward for experimental verification of their physical nature in particulars in optics and to an increasing extent in atomic and molecular physics [21-23]. In recent research of higher order EPs [24, 25] have also been attracted huge physical interest.

In this MS our aim to design an optical analogue of non-Hermitian system that hosts multiple second order EPs i.e. they corresponds only two coupling resonances. Such systems can be easily explained by introducing the following $n \times n$ block matrix.

$$H = \begin{pmatrix} H_{11} & H_{1n} \\ H_{n1} & H_{nn} \end{pmatrix} \qquad (1)$$

where each block of the Eqn. (1) should be a 2×2 elementary matrix. Each diagonal blocks explains all the essential aspects of exceptional points via coupling between two corresponding resonances. The off-diagonal blocks are intentionally set at zero because we strictly restricted the coupling between two resonances correspond to each EP whereas all other possible couplings are safely ignored. Now without any loss of generality for higher or infinite dimensional problem, just for easy illustration let consider the most general form of each diagonal block of the type $H_0^{(n)} + \lambda H_p^{(n)}$ [26] that indicates a two level quantum system $H_0^{(n)}$ with energy eigen-values $\varepsilon_1^{(n)}$ and $\varepsilon_2^{(n)}$ and the system is subjected to an external perturbation $H_p^{(n)}$ i.e.

$$H_{nn} = \begin{pmatrix} \varepsilon_1^{(n)} & 0 \\ 0 & \varepsilon_2^{(n)} \end{pmatrix} + \lambda U \begin{pmatrix} \omega_1^{(n)} & 0 \\ 0 & \omega_2^{(n)} \end{pmatrix} U^\dagger \qquad (2)$$

Where, $\quad U(\xi) = \begin{pmatrix} \cos\xi & -\sin\xi \\ \sin\xi & \cos\xi \end{pmatrix} \qquad (3)$

Here U is a unitary matrix used for similarity transformation and $\lambda$ is a real constant. $\omega_1^{(n)}$ and $\omega_1^{(n)}$ represent the coupling terms. Now the eigenvalues of H are given by

$$E_{\pm}^{n}(\lambda) = \frac{\varepsilon_1^{(n)} + \varepsilon_2^{(n)} + \lambda(\omega_1^{(n)} + \omega_2^{(n)})}{2} \pm C \quad (4)$$

Where,
$$C = \left[\left(\frac{\varepsilon_1^{(n)} - \varepsilon_2^{(n)}}{2}\right)^2 + \left(\frac{\lambda(\omega_1^{(n)} - \omega_2^{(n)})}{2}\right)^2 \right.$$
$$\left. + \frac{\lambda}{2}(\varepsilon_1^{(n)} - \varepsilon_2^{(n)})(\omega_1^{(n)} - \omega_2^{(n)})\cos 2\xi\right]^{1/2} \quad (5)$$

This model easily demonstrates the direct connection of EPs and the phenomenon of Avoided Resonance Crossing (ARC). While it is well established that interacting levels do not cross but avoid each other when the real parameter $\lambda$ is varied in this context [27]. Moreover, the ARC is observed by switching on $\xi$ that tunes the level repulsion. The two levels of the spectrum are completely degenerate when $\xi = 0$ at the point of degeneracy $\lambda_0 = (\varepsilon_1^{(n)} \quad \varepsilon_2^{(n)})/(\omega_1^{(n)} \quad \omega_2^{(n)})$ i.e. two levels of the spectrum given by, $E_{i(\xi=0)}^{(n)} = \varepsilon_i^{(n)} + \lambda \omega_i^{(n)}$, i = 1,2 intersect at $\lambda_0$. Now an exceptional point is obtained when two levels are coalesced in complex $\lambda$ plane where C should be vanished which occurs at complex conjugate point

$$\lambda_c = -\frac{\varepsilon_1^{(n)} - \varepsilon_2^{(n)}}{\omega_1^{(n)} - \omega_2^{(n)}} \exp(\pm 2i\xi) \quad (6)$$

At those points, the two levels $E_i^{(n)}(\lambda)$ coalesce and the point of coalescence is in general a square root singularity of the energy spectrum as a function of the complex parameter $\lambda$; these singularities are in general denoted exceptional points (EPs). As a consequence, it is emphasized that the use of the term coalesce is distinctly different from a degeneracy usually associated with Hermitian operators. For Hermitian operators $H_0^{(n)}$ and $H_p^{(n)}$, the EP can occur only at complex values of $\lambda$. In other words, the full problem $H_0^{(n)} + \lambda H_p^{(n)}$ is no longer Hermitian in and around the EP. At the EP the difference between degeneracy and coalescence is clearly manifested by the existence of only one eigenvector instead of the familiar two in the case of a genuine degeneracy.

The goal of this letter is to design unconventional microcavities hosting multiple EPs in its functional parameter space. Moreover, to investigate the optical performance of such cavities at operating frequencies/conditions near those discrete singular points. In this direction, first we suitably design a Fabry-Parot style optical microcavity with spatially varying gain/loss, which hosts 5 different EPs in the linear operating regime. We also observe the characteristics behavior of system eigenvalues and eigenstates in complex energy plane while encircling of one or more EPs simultaneously by a normal closed curve and closed curve with fluctuating nature.

Now, while encircling an EP, we analyze the behavior of the two-level system via characteristics behavior of the two corresponding eigenvalues by plotting the corresponding paths in complex energy plane. In the vicinity of an EP the two levels of the spectrum are strongly dependent on the interaction parameters ($\lambda$); in fact, the derivative with respect to $\lambda$ of the eigenvalues and eigenvectors is infinity at the EP i.e. it leads to a characteristic behavior of the corresponding eigenvalues under changes of the parameters. In particular, it has both neumarically and experimentally shown that an adiabatic variation of the parameters along a closed contour around an EP allows the transition from one of the resonances to the other [7, 10,14,15,17-19,21-23]. This is a characteristics feature of EP that mathematically interpreted as the consequence of the multi-valuedness of a complex variable function unveiled when going around a branch point. If a closed loop is taken in parameter space and then eigenvalues are calculated for a certain set of parameters on this loop, the eigenvalues also traverse a closed curve generally in the complex energy plane. After one cycle, the first eigenvalue will travel to the starting point of the second and vice versa which indicates that path is closed if the parameter space loop is traversed twice. As a consequence the eigenvectors of the system are altered in same way as the corresponding eigenvalues. However the eigenstates undergo an additional phase change with every additional loop i.e. signs of two eigenvectors are mutually interchanged after each circle around the EP like $\{\psi_1, \psi_2\} \to \{\psi_2, \quad \psi_1\}$. Therefore, one needs four complete turns around the EP in order to restore the original states. As a consequence, anyhow if EP is taken out from the closed curve in parameter space the eigenvalues as well as the eigenstates of the system are not permuted. They retain their respective initial state.

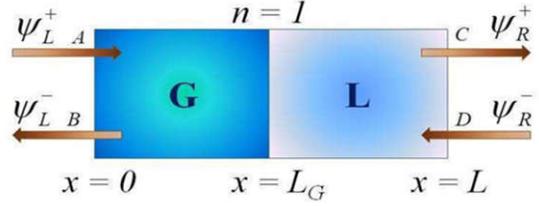

FIG.1. (Color online) Schematic diagram of a two port open interacting optical cavity that occupies the region $0 \leq x \leq L$. Gain and loss is introduced in the left and right sides of $x = L_G$. Inside the cavity, uniform background refractive index is chosen to be 1.5. A and D are the complex amplitude of incident wave corresponds to the wave function $\Psi_L^+$ and $\Psi_R^-$ whereas B and C are the complex amplitude of scattered wave corresponds to the wave function $\Psi_L^-$ and $\Psi_R^+$ respectively.

In this letter we encounter exceptional points in a specially designed Fabry-Perot type optical microcavity via formation of unconventional modes through spatially varying gain loss profile. Choice of such Fabry Perot geometry in a simplest form was driven by the fact that we can implement the output coupling along the axis, which is otherwise not possible in toroidal resonators without proper phase matching. However the results reported here are generic and is extendable to any other resonator geometries. Again, unconventional laser-absorber modes, which are created by spatially varying gain/loss profile, have been widely studied in the context of PT symmetry cavity [4, 28]. For brevity, away from the PT symmetry limit, we designed a partially pumped 1D two port open Fabry-Perot-type optical cavity, as shown in FIG. 1 [5] where A and D are the complex amplitude of incident wave whereas B and C are the complex amplitude of scattered wave from the left and the right sides of the cavity respectively. During operations, cavity is accompanied by an avoided crossing of the poles of the corresponding scattering matrix (S-matrix), which are analogous to the complex eigen-values of the associated non-Hermitian Hamiltonian of the cavity. At an essential singularity, S-matrix has produced observable resonances, and branch points in the complex energy plane. The cavity occupies the region $0 \leq x \leq L$, where the gain (denoted as co-efficient $\gamma$) is introduced in the region $0 \leq x \leq L_G$ with refractive index $n_G$ and a loss $\tau\gamma$ ($\tau$ is the fractional loss co-efficient that necessarily governs system openness and coupling strengths between the states) is introduced in the region $L_G \leq x \leq L$ with refractive index $n_L$. Thus the system is designed in such a way that if certain amount of gain is added then proportionate loss is introduced simultaneously and dictated by τ. While designing our system we choose $L$ and $L_G$ to be 10μm and 5μm respectively. Introducing $n_R$ as real part of the background refractive index, $n_L$ and $n_G$ can be written in terms of gain and loss as $n_G = n_R - i\gamma$ and $n_L = n_R - i\tau\gamma$. Here $n_R$ is chosen as 1.5 for fabrication feasibility of practical devices. Singular solutions of corresponding systems can be obtained via S-matrix formulation [4, 5, 29]. Now defining the S-matrix of such specially designed cavity as

$$\begin{bmatrix} B \\ C \end{bmatrix} = S(n(x),\omega) \begin{bmatrix} A \\ D \end{bmatrix} \quad (7)$$

we solve the equation $1/max[eig\ S(\omega)] = 0$ by using numerical root finding method to find the matrix poles. Here $max[eig\ S(\omega)]$ is the maximal-modulus eigenvalues of the matrix $S(\omega)$. The poles occur only at complex values of $k$ in the lower half plane i.e. with a negative imaginary part. Now, we deliberately find a pair of poles of S-matrix with in the lower half of the complex frequency plane for a certain frequency limit. The pair is coupled after introducing certain amount of gain/loss. Accordingly, when non-uniform gain is added to each pair individually then chosen poles of that pair are mutually coupled. With the increasing of gain up to certain fixed amount, only by proper tuning over τ we can observe two different type of the state repulsion phenomenon with crossing of real parts with simultaneous avoided crossing of corresponding imaginary parts of the poles or vice versa in the eigenvalue plot of the cavity. As a consequence, while tuning over τ for a certain range of γ, one should find a point $(\gamma_{EP}, \tau_{EP})$ in $(\gamma, \tau)$ plane that appears as an exceptional point where two levels corresponds to the chosen pair of poles collapse. Accordingly with an increasing gain up to 0.1 and suitable tuning over τ, we have embedded five different EPs corresponds to the five different pairs of poles.

TABLE 1: Examples of exceptional points in (γ, τ) plane with their real and imaginary part of the energies. All values are given in atomic units. The numbers are used as labels to identify the exceptional points.

| EPs | γ | τ | ~Re[k] | ~Im[k] |
|---|---|---|---|---|
| 1 | 0.0944 | 0.11845 | 7.6437327166 | 0.0998070443 |
| 2 | 0.0902 | 0.12210 | 8.0627233175 | 0.0990508991 |
| 3 | 0.0863 | 0.12585 | 8.4815449508 | 0.1004088978 |
| 4 | 0.0828 | 0.12905 | 8.9004991875 | 0.1004605122 |
| 5 | 0.0796 | 0.13205 | 9.3194540881 | 0.1002486813 |

They are listed in TABLE 1 with corresponding approximate real and imaginary part of the energy values. In FIG. 2 (a) we plot all these EPs to show their linear appearance in parameter plane. Here, the fitted curve, marked by blue dotted line, indicates the appearance of them along a straight line, denoted as Branch Line in (γ, τ) plane with tangent -0.9225 a.u. and intercept 0.2054 a.u. from positive τ axis. Due to appearance of EPs in a single branch line we can shift from one state to another simply by switching the depending parameters. Now, while enclosing the EP in (γ, τ) plane, one good choice for enclosing path is circle with certain characteristics parameter $a < 1$ and relative to the center at corresponding EP. Hence in parameter plane the closed loops are followed by the parametric equations of φ (goes from 0 to 2π) as given by

$$\gamma(\phi) = \gamma_{EP}(1 + a\cos\phi) \quad \text{and} \quad \tau(\phi) = \tau_{EP}(1 + a\sin\phi) \quad (8)$$

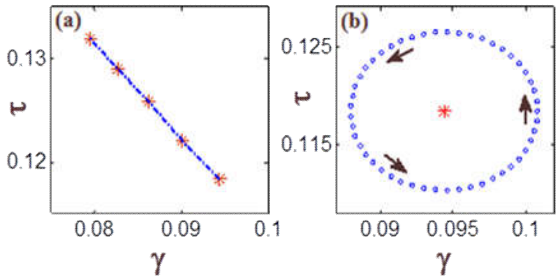

FIG.2. (Color online) (a) Five nearest EPs in parameter plane for our designed cavity marked by five red stars. Blue dotted line is the fitting of five EPs that indicates the appearance of them along a branch line. (b) Encircling of one EP followed by equation (8) with characteristics parameter a = 0.04 a.u. Arrows indicate the direction of progression.

In FIG. 2(b) we show the enclosing of one EP followed by equation (8) with $a$ as 0.04 (a.u). Arrows indicate the direction of progration.

Now we analyze the system rigidity via the characteristics behavior of eigenvalues and eigenstates while encircling an EP. To serve this purpose, we randomly choose three nearest EP from TABLE 1. Since for a system EPs appear along a branch line, simply by varying '$a$' we could fix the number of EPs inside one closed circle. FIGs. 3-6 show typical numerical results in complex k-plane. In FIG. 3, each point of the trajectory represents one of the eigenvalues for a point of the enclosing circle in parameter plane shown at inset of FIG. 3 corresponds to a set of γ and τ, whereas two red stars at the center of two blue circles denotes the approximate positions of $EP_1$ and $EP_2$. The characteristics parameter of the circle is set at 0.04 a.u. according to equation (8) such that each circle encloses only the corresponding EP. As a consequence of EP being the second order branch point in eigenvalues, after going one round enclosing the each EP, the corresponding first eigenvalue moves to the starting point of the second eigenvalues (marked by dotted red line) and after next round vice versa (marked by dotted blue line) and then form a complete closed loop. The arrows indicate the direction of progression.

Now to show the mutual dependence between two consecutive EPs, in FIG. 4(a) and 4(b), we demonstrate the characteristics behavior of eigenvalues correspond to $EP_2$ (marked by red star in both inset of FIG. 4) for its two different positions followed by the enclosing parameter values with center at $EP_1$ (marked by blue star in both inset of FIG. 4). In FIG. 4(a) following the blue circular path, that encircle $EP_1$ but not $EP_2$ with characteristics parameter $a = 0.04$ a. u. (described in inset) for each value of $(\gamma, \tau)$ we illustrate the trajectory of two eigenvalues correspond to $EP_2$ by plotting them in complex k-plane. Here after going two round enclosing the $EP_1$ in $(\gamma, \tau)$ plane both eigenvalues corresponding to $EP_2$ makes individual loop in complex k-plane without any permutation (shown by dotted red and blue line respectively). On the other hand in FIG. 4(b) we increase $a$ to 0.065 a.u i.e. in this case both $EP_1$ and $EP_2$ are being encircled simultaneously. Now for each value of $(\gamma, \tau)$ on the circular path illustrated at inset, two eigenvalues correspond to $EP_2$ are being permuted in similar way demonstrated in FIG. 3.

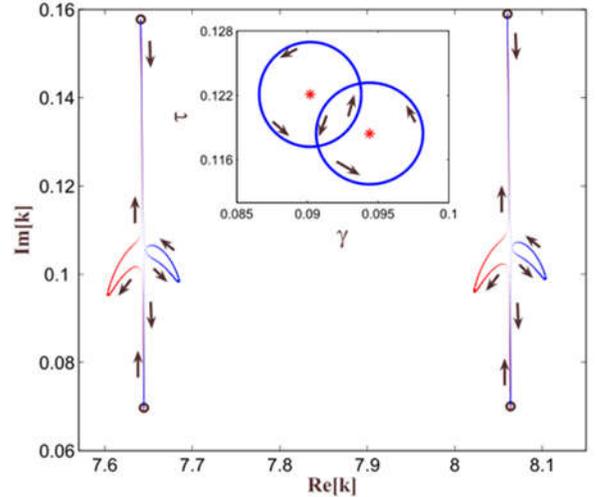

FIG.3. (Color online) Trajectories of the eigenvalues in complex energy plane correspond to two EPs followed by the path in (γ, τ) plane as described at the inset where blue lines indicates the enclosing circular paths with respect to the center at corresponding exceptional points marked by two red stars and with $a = 0.04\ a.u.$ i.e. each path enclose only one EP. The arrows indicate the direction of propagation. In complex k-plane path marked by dotted red and blue line correspond to the first and second round encircling in parameter plane respectively.

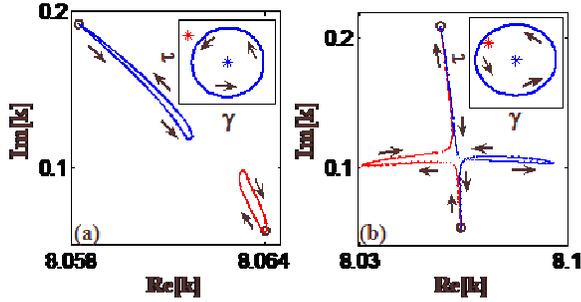

FIG.4. (Color online) Trajectories of the eigenvalues in complex energy plane correspond to $EP_1$ (marked by red star at both inset) followed by the path in $(\gamma, \tau)$ plane as described by the blue circle with respect to the center at $EP_2$ (marked by blue star) and (a) with $a = 0.04\ a.u.$ i.e. the path doesn't enclose $EP_1$; (b) with $a = 0.065\ a.u.$ i.e. the path encloses $EP_1$. In complex k-plane the eigenvalues traverse a closed curve individually marked by dotted red and blue line. The arrows indicate the direction of progression.

In FIG. 5 we show the characteristics behavior of eigenvalues in complex k-plane corresponding to three nearest EP (marked by three red stars at inset) followed by the particular encircling parameter values in parameter plane. To perform this we consider an enclosing parameter circle followed by equation (8) where the center of the circle $(\gamma_{EP}, \tau_{EP})$ in equation (8) is replaced by any arbitrary center, say at (0.08825, 0.12395), i.e. none of the EP presents the center of the circle. We set characteristics parameter for the enclosing circle in parameter space at 0.068 a.u such that only $EP_2$ and $EP_3$ are enclosed by the circle whereas $EP_1$ stays at the outside of the circle. Now after successfully encircling two rounds in parameter space (shown by blue line in inset), the eigenvalues corresponding to $EP_2$ and $EP_3$ are permuted and form a complete loop in complex k-plane whereas eigenvalues corresponding to $EP_2$ make individual loop without any permutation (shown by dotted red and blue line respectively). Thus there is no effect on the eigenvalues trajectories correspond to one EP due to the presence of another EP inside the closed loop in parameter plane.

In above cases we study the characteristics behavior of system eigenvalues in complex k-plane and their mutual effects while the corresponding EPs are enclosed by a simple closed curve. Analyzing these results it can be concluded that if the EP is inside the loop then the eigenvalues of two corresponding states being transformed into each other and all other states are returned themselves at the end of the loop (shown in FIG. 3 and FIG. 5). Consequently if the EP is taken out from the loop anyhow then the corresponding eigenstates retain in same states without any permutation on corresponding eigenvalues (shown in FIG. 4 and FIG. 5). FIG. 4 also indicates that there are no effects on the state correspond to one EP by the enclosing parameters of another EP. This mutual independency of the EP corresponding to a set of different states represents a good evidence about system rigidity.

Now we illustrate the system eigenvalues behavior in complex k-plane while the corresponding EP is enclosed by closed curve with random fluctuation. In this direction we enclose the $EP_1$ (marked by red star) by a randomly fluctuated circular path with center at corresponding EP (characteristics parameter a= 0.04 a.u.) shown by blue line in the inset of FIG. 6 and then traverse trajectory for each set of (γ, τ). FIG. 6 illustrated the trajectories of the two corresponding eigenvalues in complex k-plane. Here also after going one round encircling the EP along fluctuated curve in 2D parameter plane, the first eigenvalue travels to starting point of the another one and vice versa (shown by the trajectories of red and blue dotted lines). Consequently, two complete round enclosing EP is required to form a closed loop in complex k-plane. This result is also similar to the results shown in FIG. 4. This result indicates that there is no effect of the fluctuations along the enclosed path on the behavior of system eigenvalues.

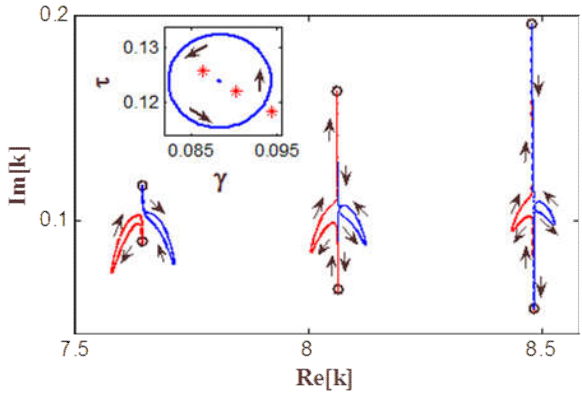

FIG.5. (Color online) Trajectories of the eigenvalues in complex energy plane correspond to three EPs (marked by three red stars at inset) followed by the path in $(\gamma, \tau)$ plane as described at the inset where blue lines indicate the enclosing circular paths with respect to an arbitrary center marked by blue dot at (0.08825, 0.12395) with $a = 0.068\ a.u.$ i.e. the path encloses only two nearest EP whereas one EP at outside the closed path. The arrows indicate the direction of propagation. In complex k-plane path marked by dotted red and blue line correspond to the first and second round encircling in parameter plane respectively.

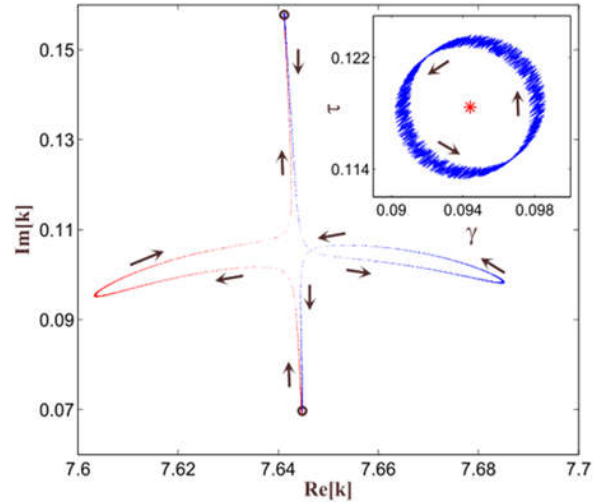

FIG.6. (Color online) Trajectories of the eigenvalues in complex energy plane corresponding to $EP_2$ followed by the fluctuating path in $(\gamma, \tau)$ plane as described at the inset where blue lines indicates the enclosing circular path with random fluctuations with respect to the center at corresponding exceptional point marked by red star and with $a = 0.04\ a.u.$ The arrows indicate the direction of progression. In complex k-plane path marked by dotted red and blue line correspond to the first and second round enclosing in parameter plane respectively.

In summary, beyond PT symmetry limit, numerically, we have found the exceptional points in a Fabry-Perot styled optical microcavity with spatially varying gain/loss. For a particular system EPs are appeared along branch line. While encircling an exceptional point in parameter space, if the EP is inside the loop then the eigenvalues of two corresponding states have been permuted. Otherwise, it is found that the eigenvalues of individual mode also traverse a closed curve. We use these characteristics behavior of system eigenvalues to analyze the rigidity of the system under various conditions. We have shown the interesting dynamics of eigenvalues depending only on position of the corresponding EP with respect to the enclosing path i.e. whether it is inside the enclosed loop or not. By adding deliberate fluctuations we have shown that system eigenvalues' characteristics behaviors are also irrespective of the enclosed path around EP. By showing these mutual independency of the parameters correspond to the EPs and enclosing path independency on the system eigenvalues and eigenstates, we demonstrate the stability and rigidity of a system which open up a better platform to design high performance photonic devices based on such unconventional microcavities.

**Funding sources and acknowledgments:** S. Ghosh acknowledges the financial support by Department of Science and Technology, India as a INSPIRE Faculty Fellow [IFA-12; PH-13].